
\magnification=1200
\def\gl{\mathrel{\raise1ex\hbox{$>$\kern-.75em\lower1ex\hbox{$<$}}}}
\def\lg{\mathrel{\raise1ex\hbox{$<$\kern-.75em\lower1ex\hbox{$>$}}}}
\def\gtwid{\mathrel{\raise.3ex\hbox{$>$\kern-.75em\lower1ex\hbox{$\sim$}}}}
\def\ltwid{\mathrel{\raise.3ex\hbox{$<$\kern-.75em\lower1ex\hbox{$\sim$}}}}
\def\sqr#1#2{{\vcenter{\hrule height.#2pt
      \hbox{\vrule width.#2pt height#1pt \kern#1pt
         \vrule width.#2pt}
      \hrule height.#2pt}}}
\def\square{\mathchoice\sqr34\sqr34\sqr{2.1}3\sqr{1.5}3}

\def\ie{\hbox{{\it i.\ e.}}}\def\etc{{\it etc.}}
\def\ea{\hbox{{\it et al.}}}


\def\leaderfill{\leaders\hbox to 1em{\hss.\hss}\hfill}


\def\CF{\hbox{{$\cal F$}}}

\def\CO{\hbox{{$\cal O$}}}

\def\ref#1{${}^{#1}$}

\newcount\eqnum \eqnum=0  
\newcount\eqnA\eqnA=0\newcount\eqnB\eqnB=0\newcount\eqnC\eqnC=0
\newcount\eqnD\eqnD=0
\def\eqnoi{\global\advance\eqnum by 1\eqno(\the\eqnum)}
\def\eqnai{\global\advance\eqnum by 1\eqno(\the\eqnum{a})}
\def\eqnoA{\global\advance\eqnA by 1\eqno(A\the\eqnA)}
\def\eqnoB{\global\advance\eqnB by 1\eqno(B\the\eqnB)}
\def\eqnoC{\global\advance\eqnC by 1\eqno(C\the\eqnC)}
\def\eqnoD{\global\advance\eqnD by 1\eqno(D\the\eqnD)}
\def\back#1{{\advance\eqnum by-#1 Eq.~(\the\eqnum)}}
\def\backs#1{{\advance\eqnum by-#1 Eqs.~(\the\eqnum)}}
\def\backn#1{{\advance\eqnum by-#1 (\the\eqnum)}}
\def\backA#1{{\advance\eqnA by-#1 Eq.~(A\the\eqnA)}}
\def\backB#1{{\advance\eqnB by-#1 Eq.~(B\the\eqnB)}}
\def\backC#1{{\advance\eqnC by-#1 Eq.~(C\the\eqnC)}}
\def\backD#1{{\advance\eqnD by-#1 Eq.~(D\the\eqnD)}}
\def\last{Eq.~(\the\eqnum)}                   
\def\lasts{Eqs.~(\the\eqnum)}                   
\def\lastn{(\the\eqnum)}                      
\def\lastA{Eq.~(A\the\eqnA)}\def\lastB{Eq.~(B\the\eqnB)}
\def\lastC{Eq.~(C\the\eqnC)}\def\lastD{Eq.~(D\the\eqnD)}
\newcount\refnum\refnum=0  
\def\refi{\smallskip\global\advance\refnum by 1\item{\the\refnum.}}

\newcount\rfignum\rfignum=0  
\def\rfigi{\medskip\global\advance\rfignum by 1\item{Figure \the\rfignum.}}

\newcount\fignum\fignum=0  
\def\figi{\global\advance\fignum by 1 Fig.~\the\fignum}

\newcount\secnum\secnum=0 
\def\chap#1{\global\advance\secnum by 1
\bigskip\centerline{\bf{\the\secnum}. #1}\smallskip\noindent}
\def\subsec#1{\smallskip{\bf\centerline{#1}\smallskip}}
\def\tsubsec#1{{\bf\centerline{#1}\smallskip}}

\def\p2d#1#2{{\partial^2 #1\over\partial #2^2}} 
\def\td#1#2{{d #1\over d #2}}      
\def\t2d#1#2{{d^2 #1\over d #2^2}} 
\def\av#1{\langle #1\rangle}                    

\def\th{{${}^{\rm th}$}}

\def\jth{{$j^{\rm th}$}}
\def\2kth{{$2k^{\rm th}$}}

\def\nth{{$n^{\rm th}$}}

\def\n-th{{$(n-1)^{\rm th}$}}

\def\N-th{{$(N-1)^{\rm th}$}}

\def\0th{$0^{\rm th}$}
\def\1st{$1^{\rm st}$}
\def\2nd{$2^{\rm nd}$}
\def\3rd{$3^{\rm rd}$}
\def\4th{$4^{\rm th}$}
\def\5th{$5^{\rm th}$}
\def\5th{$6^{\rm th}$}
\def\6th{$7^{\rm th}$}
\def\7th{$7^{\rm th}$}
\def\8th{$8^{\rm th}$}
\def\9th{$9^{\rm th}$}


\def\a{{\alpha}}
\def\b{{\beta}}
\def\G{\Gamma}
\def\D{\Delta}
\def\e{\epsilon}

\def\l{\lambda}
\def\m{\mu}
\def\n{\nu}
\def\O{\Omega}
\def\p{\pi}

\def\t{\tau}
\def\th{\theta}

\overfullrule=0pt

\def\intt{\int _0 ^{\pi} d\th\, (\sin\th)^{d-2}}

\def\intte{\int _0 ^{\pi} d\th\, (\sin\th)^{d-2}\sqrt{1+\e^2+2\e\cos\th}}

\def\Cax{{\cal C}_0(\xi)}
\def\Cbx{{\cal C}_{\pm}(\xi)}
\def\Cat{{\cal C}_0(t\D^3)}
\def\Cbt{{\cal C}_{\pm}(t\D^3)}


\centerline{\bf BALLISTIC ANNIHILATION KINETICS:}
\centerline{\bf THE CASE OF DISCRETE VELOCITY DISTRIBUTIONS}
\bigskip
\bigskip
\centerline{\bf P.~L.~Krapivsky and S.~Redner}
\smallskip
\centerline{Center for Polymer Studies and Department of Physics}
\centerline{Boston University, Boston, MA 02215, USA}
\bigskip
\centerline{\bf F.~Leyvraz}
\smallskip
\centerline{Instituto de F\'\i sica---UNAM, Laboratorio de Cuernavaca}
\centerline{Apdo.~postal 20--364, 01000 M\'exico D. F., MEXICO}
\vskip 0.3in
\centerline{\bf ABSTRACT}
\smallskip
{\rightskip=3truepc \leftskip=3truepc
\baselineskip=12truept\noindent

The kinetics of the annihilation process, $A+A\to 0$, with ballistic
particle motion is investigated when the distribution of particle
velocities is {\it discrete}.  This discreteness is the source of many
intriguing phenomena.  In the mean field limit, the densities of
different velocity species decay in time with different power law rates
for many initial conditions.  For a one-dimensional symmetric system
containing particles with velocity 0 and $\pm 1$, there is a particular
initial state for which the concentrations of all three species as decay
as $t^{-2/3}$.  For the case of a fast ``impurity'' in a symmetric
background of $+$ and $-$ particles, the impurity survival probability
decays as $\exp(-{\rm const.}\times \ln^2t)$.  In a symmetric 4-velocity
system in which there are particles with velocities $\pm v_1$ and $\pm
v_2$, there again is a special initial condition where the two species
decay at the same rate, $t^{-\a}$, with $\a\cong 0.72$.  Efficient
algorithms are introduced to perform the large-scale simulations
necessary to observe these unusual phenomena clearly.

\vglue 0.6truecm}

\vfil
\chap{INTRODUCTION}

In this article, we describe some intriguing aspects of the reaction
kinetics in single species annihilation, $A+A\to 0$, when particles move
ballistically with a {\it discrete} distribution of velocities.
Unexpected long-time phenomena occur which depend fundamentally on the
form of the initial velocity distribution.  The results discussed here
are complementary to our earlier work on ballistic annihilation with a
continuous distribution of particle velocities [1].  For this latter
system, the exponents characterizing the decay of the concentration and
the typical velocity depend continuously on the form of the initial
velocity distribution.  For discrete velocity distributions, however,
the decay kinetics exhibits a fundamentally richer character with
fundamental differences in long-time behavior for small changes in the
initial conditions.

Our investigation is also inspired by earlier work by Elskens and
Frisch and independently Krug and Spohn who considered the kinetics
of the ``2-velocity'', or ``$\pm$'' model in one dimension [2].
Here, the initial velocity distribution of reactants is
$$
P(v,t=0)=p_+\delta(v-v_0)+p_-\delta(v+v_0),
$$
with $p_++p_-=1$.  The spatial distribution of reactants has minimal
influence on the kinetics as long as the distribution is non-singular.
For convenience, we therefore consider the distribution to be Poisson in
this paper.  When $p_+>p_-$, the majority species quickly reaches a
finite asymptotic limit, while the minority density decays exponentially
in time.  In the interesting situation where the initial densities of
the two species are equal, the density decays as [2]
$$
c(t)\propto \sqrt{c(0)/v_0t}\eqnoi
$$
in the long time limit.  This relatively slow decay, compared to the
rate equation prediction of $t^{-1}$ stems from initial density
fluctuations.  In a region of length $L$ there will typically be an
imbalance of the order of $\sqrt{L}$ in the number of + and $-$
particles.  After a time $t\cong L$ has elapsed, only this initial
number difference will remain within the region.  Therefore the local
particle number is proportional to $\sqrt{L}$, and \last\ follows.  Thus
the system organizes into domains of like velocity particles whose
typical size grows linearly in time as the reaction proceeds (Fig.~1).

Consider now a simple and natural generalization to the ``3-velocity''
model [3].  Without loss of generality, the initial distribution of
velocities may be written as
$$
P(v,t=0)=p_+\delta(v-v_+)+p_-\delta(v+1)+ p_0\delta(v),\eqnoi
$$
with $p_++p_-+p_0=1$.  We will primarily focus on the symmetric case
where $v_+=1$ and $p_+=p_-=p_\pm$.  The space-time evolution of this
system in one dimension for two representative values of $(p_\pm,p_0)$
is shown in Fig.~1.  One of our basic goals is to understand the time
dependence of the mobile and stationary concentrations for different
initial conditions.  Particularly intriguing is the transition from a
regime where the stationary particles persist, for $p_0>1/4$, to a
regime where stationary particles decay more rapidly than the mobile
particles, for $p_0<1/4$.  At a ``tricritical'' point located at
$p_0=1/4$, the concentrations of both the mobile and stationary species
decay as $t^{-2/3}$ [4].  While there is now a theoretical approach to
compute this exponent exactly [5], there is not yet an intuitive
understanding of this striking behavior.  Another intriguing facet of
this system is the decay of a ``fast impurity'', namely, a single
particle with velocity $+1$ in system with equal concentrations of 0 and
$-1$ particles.  By considering the dominant contributions to the
impurity survival probability, we find that this quantity decays
asymptotically as $\exp(-{\rm const.}\times\ln^2t)$.

Another class of interesting behavior is exemplified by the
symmetric 4-velocity model with initial velocity distribution
$$
P(v,t=0)=p_1(\delta(v-v_1)+\delta(v+v_1))+
p_2(\delta(v-v_2)+\delta(v+v_2)),
\eqnoi
$$
with $v_2>v_1$ and $p_1+p_2=1$.  According to the rate equations, the
more mobile species decays as $t^{-v_2/v_1}$ while the less mobile
decays as $t^{-1}$, independent of the initial concentrations.  In one
dimension, however, either the slower or the faster particles dominate
in the long time limit, depending on their relative initial
concentrations.  At a critical value of $p_1/p_2$ which depends on
$v_1/v_2$, numerical simulations indicate that both species decay as
$t^{-\a}$, with $\a\cong 0.72$.  We shall also argue that systems with
symmetric velocity distributions with $n>4$ components exhibit behavior
which is characteristic of either the 3-velocity model, if $n$ is odd,
or 4-velocity model, if $n$ is even.  Thus we focus primarily on the 3-
and 4-velocity models as the simplest in the family of symmetric
discrete velocity models.

In the next two sections, we discuss the annihilation kinetics of
discrete velocity models in the mean-field limit.  We first treat the
conventional rate equations which are based on one-dimensional
kinematics.  The shortcoming inherent in the assumption naturally leads
us to consider ``constant speed'' models and the correct $d$-dimensional
kinematics.  We find rich kinetic behavior which depends on the ratios
of initial concentrations, the particle radii, and the speeds of the
different species.  In section 4, we study the kinetics of
one-dimensional systems.  Given the subtle nature of many of our
observations, relatively efficient and specialized algorithms were
developed to provide sufficient data to determine the long-time behavior
with confidence.  For the symmetric 3-velocity model, we investigate the
$t^{-2/3}$ decay associated with the tricritical point and the
$\exp(-{\rm const.}\times\ln^2t)$ decay of the fast impurity problem, in
detail.  We then consider the kinetics of the symmetric 4-velocity
model, once again concentrating on the multicritical behavior associated
with the initial condition where both the fast and slow particles decay
at the same rate.

\chap{MEAN FIELD THEORY WITH ONE DIMENSIONAL KINEMATICS}
\tsubsec{(a) The 3-Velocity Model}

The mean-field rate equations for the 3-velocity model are deceptively
simple, but lead to relatively complex behavior.  For symmetric velocity
distribution, the rate equations for the concentrations of the
left-moving, right-moving, and stationary species, $c_-(t)$, $c_+(t)$,
and $c_0(t)$, respectively, are
$$
\eqalign{
\dot c_0 &= -c_0(c_++c_-),\cr
\dot c_+  &= -c_+(c_0 + 2c_-),\cr
\dot c_-  &= -c_-(c_0 + 2c_+),\cr}
\eqnoi
$$
where the overdot denotes time derivative.  The numerical factors of 2
reflect the fact that the rate of a $+\,-$ collision is twice that of
$+\,0$ or $-\,0$ collisions, if we assume that particles move only in
one dimension.  It is in this spirit that the above rate equations are
referred to as mean-field theory with one-dimensional kinematics.  A
more complete approach which incorporates
$d$-dimensional kinematics will be outlined in the next section.

To solve these equations, it is helpful to rewrite the rate equations in
terms of $\psi\equiv (c_++c_-)/c_0$ and $\phi\equiv(c_+-c_-)/c_0$, and
the modified time $dx\equiv c_0\,dt$.  This gives,
$$
\eqalign{
\psi'+\psi&=\phi^2,\cr
\phi'+\phi&=\phi\psi, \cr}
\eqnoi
$$
where the prime denotes differentiation with respect to $x$.
Use of the integrating factors $\Psi=\psi\,e^x$ and
$\Phi=\phi\,e^x$ simplify these equations to
$$
\eqalign{
\Psi'&=\Phi^2\,e^{-x},\cr
\Phi'&=\Phi\Psi\,e^{-x}\cr},
\eqnoi
$$
from which it is evident that $\Psi^2-\Phi^2={\rm const.}\equiv a^2>0$.
Thus the equation of motion for $\Psi$ becomes $\Psi'=
e^{-x}(\Psi^2-a^2)$, with solution
$$
\Psi(y)=a\coth\left[\coth^{-1}\left({\Psi(0)\over a}\right) - ay\right],
\eqnoi
$$
where he have introduced the new time-like variable $dy=e^{-x}\,dx$,
with $y=1-e^{-x}$ a monotone increasing function of $t$.

To classify the long-time behavior, consider the relative composition
triangle $p_++p_-+p_0=1$.  As indicated in Fig.~2, a given initial
condition typically evolves to a ``phase'' where only a single species
persists in the long time limit.  Consider first the stationary, or
``0'', phase, $c_0(\infty)>0$.  From the definitions of $x$ and $y$, the
condition $c_0(\infty)>0$ implies that $y\sim 1-e^{-at}$ as
$t\to\infty$, which leads to $\psi\sim e^{-at}$.  Thus the
concentrations of the mobile species decay exponentially in time in the
0 phase.  Interestingly, for $p_+=p_-$ stationary particles persist even
if the initial concentration of stationary particles becomes small.
However, the width of this phase becomes vanishingly small in this
limit.  To determine this width, note, from \back3, that on the boundary
where the densities of the stationary and positive particles decay at
the same rate, the asymptotic solutions of the rate equations are $c_+,
c_0 \propto 1/t$ while $c_- \sim 1/t^3$.  Thus the boundary between the
0 and + phases can be identified by the ratio $\psi$ approaching a
finite limit as $t\to\infty$.  Since $1-y$ approaches 0 as $t\to\infty$,
a finite limiting value for $\psi=e^{-x}\Psi$ requires that the
argument of the hyperbolic cotangent in \last\ goes to zero.  One
thereby finds that the width of the 0 phase region vanishes as
$\exp(-1/c_0(0))$ as $c_0(0)\to 0$.

For the symmetric system, a detailed computation gives the asymptotics
$$
\eqalign{
c_\pm(t) &\sim{1\over 2}c_0(\infty)
G(\l)\,e^{-c_0(\infty)t},\cr
c_0(t)   &\sim c_0(\infty)
\exp\left[G(\l)e^{-c_0(\infty)t}\right],\cr}
\eqnoi
$$
with $G(\l)\equiv\exp\left[\int_1^{\l}{dz\over
z}e^{-z}-\int_0^1{dz\over z}\left(1-e^{-z}\right)\right]$,
$\l \equiv 2c_\pm(0)/c_0(0)$, and with the final density of stationary
particles given by
$$
c_0(\infty)=c_0(0)\,\,e^{-2c_\pm(0)/c_0(0)}.
\eqnoi
$$
Thus while a residue of stationary particles always persists, their
concentration is astronomically small if the initial
concentration of stationaries is small.

For the asymmetric 3-velocity model, the corresponding rate
equations are
$$
\eqalign{
\dot c_0 &= -c_0(vc_+ + c_-),\cr
\dot c_+  &= -c_+(vc_0 + (1+v)c_-),\cr
\dot c_-  &= -c_-(c_0 + (1+v)c_+).\cr}
\eqnoi
$$
While we are unable to solve these equations, we can readily find the
asymptotic behavior.  In the 0 phase, the rate equations for the mobile
particles have the asymptotic solutions $c_+(t)\sim e^{-tvc_0(\infty)}$ and
$c_-(t)\sim e^{-tc_0(\infty)}$.  Similar exponential decays
characterize the behavior of the minority species in the other two
phases.  On the separatrices, however, two species decay at the same
rate while the minority species decays faster. For example, on the
separatrix between the 0 and + phases, one has $c_0\sim c_+\gg c_-$.
Substituting this into \last\ yields the asymptotic solution $c_+, c_0
\simeq 1/vt$ while $c_- \sim t^{-1-2/v}$. Similarly, on the separatrix
between the 0 and $-$ phases $c_-, c_0 \simeq 1/t$ and $c_+ \sim
t^{-1-2v}$.

The long-time persistence of stationary particles also occurs
in a general $2m+1$-component model with velocities
$0=v_0<v_1<\ldots<v_m$ and corresponding concentrations
$c_0(t), c_1(t), \ldots, c_m(t)$. The rate equations for these
concentrations are
$$
\eqalign{
\dot c_0 &= -2c_0\sum_{j=1}^m v_jc_j,\cr
\dot c_k  &= -c_0c_kv_k
-2c_k\left(v_k\sum_{j=1}^{k-1} c_j + \sum_{j=k}^m v_jc_j \right).\cr}
\eqnoi
$$
Introducing $x=\int_0^tdt'\,c_1(t')$ and the dimensionless
concentrations $\phi_k=c_k/c_0$, we
obtain a closed system of equations for $\phi_k(x)$
and an additional equation for $c_0(x)$
$$
\eqalign{\td {\ln\phi_k} x  &= -v_k - 2\sum_{j=1}^{k-1}(v_k - v_j)\phi_j,
\quad k=1,\ldots, m\cr
\td {\ln c_0} x  &= - 2\sum_{j=1}^{m} v_j\phi_j.\cr}
\eqnoi
$$
Since $v_k>v_j$ for $k>j$, the first of \lasts\ gives $\phi_k(x) \le
\phi_k(0)e^{-v_kx}$.  Substituting these  into the equation for $c_0$ yields
$$
\ln {c_0(x)\over c_0(0)} \ge - 2\sum_{j=1}^{m}\phi_j(0)(1-e^{-v_jx})
\eqnoi
$$
which immediately leads to the lower bound for final density
of stationary particles
$$
c_0(\infty) \ge c_0(0)\exp\left[- {2\over c_0(0)}\sum_{j=1}^{m}c_j(0)\right].
\eqnoi
$$
For the 3-velocity case this bound is exact and coincides with \back5.
Thus stationary particles always survive in the symmetric
$(2m+1)$-velocity model, although the final residue is vanishingly small
when their initial concentration is small.
\vfill\eject
\subsec{(b) The 4-Velocity Model}

For the symmetric 4-velocity model, denote by $c_1$ and $c_2$ the
concentrations of species with velocities $\pm v_1$ and $\pm
v_2$, respectively.  Without loss of generality, let $v_2>v_1$ and
set $v_1=1, v_2=v$.  The rate equations for this system are,
$$
\eqalign{
\dot c_1 &= -2c_1^2 -2vc_1c_2,\cr
\dot c_2 &= -2vc_2^2 -2vc_1c_2,}\eqnoi
$$
with asymptotic solution
$$
c_1(t)\sim t^{-1}, \quad c_2(t)\sim t^{-v}.
\eqnoi
$$
Thus the faster species decays non-universally as $t^{-v}$.  This
asymptotic behavior is reached only at very long times, however, when
the initial velocities are nearly identical.  Essentially the same
equations were solved in the context of heterogeneous diffusive
single-species annihilation in which particles have different diffusion
coefficients [6].

Analogous behavior occurs in the symmetric $2m$-velocity model with
concentrations $c_1(t), \ldots, c_m(t)$ and speeds $\pm v_j$, with
$v_1<\ldots<v_m$.  The rate equations are
$$
\dot c_k=-2c_k\left(v_k\sum_{j=1}^{k-1} c_j + \sum_{j=k}^m v_jc_j \right).
\eqnoi
$$
Introducing now $x=2\int_0^tdt'\,c_1(t')$ and
$\phi_k=c_k/c_1, \phi_1 \equiv 1$, we obtain
$$
\eqalign{
\td {\ln\phi_k} x  &=-\sum_{j=1}^{k-1}(v_k-v_j)\phi_j,
\quad 2\le k \le m,\cr
\td {\ln c_1} x &= - \sum_{j=1}^{m} v_j\phi_j.\cr}
\eqnoi
$$
As in the $(2m+1)$-velocity model, one can straightforwardly derive
$\phi_k(x) \le \phi_k(0)e^{-(v_k-v_1)x}$.  This, together with the
equation for $c_1$
and the relation $\phi_1 \equiv 1$, show that $c_1 \sim e^{-v_1x}$.
Combining this result with the definition of $x$ proves that
$x\to\infty$ as $t\to\infty$.  It therefore follows that in the long
time limit $\phi_k(x) \sim e^{-(v_k-v_1)x}$.
Re-expressing this in terms of $c_j(t)$ leads to
$$
c_1 \sim t^{-1}, \quad
c_2 \sim t^{-v_2/v_1}, \quad \ldots, \quad
c_m \sim t^{-v_m/v_1}.
\eqnoi
$$
Thus a more mobile species $k$ decays non-universally with an associated
exponent equal to the velocity ratio $v_k/v_1$.

\chap{MEAN FIELD THEORY WITH $D$-DIMENSIONAL KINEMATICS}

We now generalize the rate equations to account for $d$-dimensional
kinematics.  This should be viewed as the ``true'' mean-field theory of
ballistic annihilation.  It is convenient to consider first the kinetics
of a single impurity in a background of particles moving at the same
speed.  From this, the general mean-field theory follows naturally.

\subsec{(a) The Impurity Limit}

Let background particles of radii $R$ move with identical velocities
$\bf v$ which are uniformly distributed in angle, \ie, $P({\bf
v},t=0)=\delta(|{\bf v}|-v_0)$.  If we temporarily neglect the
annihilation events among background particles, then by an elementary
mean-free path argument, the (infinitesimal) concentration of the
impurity species with radius $R_I$ varies as
$$
\dot c_I = -c_Ic\,v\,\O_{d-1}(R+R_I),
\eqnoi
$$
where $\O_{d-1}(R)$ is the volume of a sphere of radius $R$ in
$d-1$ dimensions.

If the impurity moves with velocity $\bf w$, then the decay rate must
be averaged over all directions of relative velocities, $\bf v - \bf
w$.  This leads to
$$
\eqalign{
\dot c_I &= -c_Ic\times v\O_{d-1}(R+R_I)\times{\intte \over \intt},\cr
         &\equiv -c_Ic\times v\O_{d-1}(R+R_I)\,\times\,\CF(\e).\cr}
\eqnoi
$$
with $\e=w/v$.
To find the impurity concentration $c_I$ we must first determine the
background concentration $c$.  Since a
background particle can be considered as an impurity of radius
$R$ moving with velocity $v$, we apply \last\ to obtain
$\dot c = -c^2\,v\,\O_{d-1}(2R)\,\CF(1)$,
with asymptotic solution $c(t) \sim [v\O_{d-1}(2R) \CF(1) t]^{-1}$.
Using this in \last\ gives
$$
c_I\sim t^{-\a}, \qquad {\rm with\ \ }
\a=\a(R_I,\e)={\O_{d-1}(R+R_I)\,\CF(\e)\over{\O_{d-1}(2R)\,\CF(1)}}.
\eqnoi
$$
For example,
when there is a stationary particle in a uniform background of moving
particles, $c_I\sim t^{-\a_0(d)}$, with
$$
\a_0(d)=\CF(0)/\CF(1)= {\G({1\over 2})\G(d-{1\over 2})\over
2^{d-1}\G^2({d\over 2})},\eqnoi
$$
and $\G$ is the gamma function.  Note $\a_0(d)$ is rational for odd
dimensions and transcendental for even dimensions: $\a_0(1)=1,
{}~\a_0(2)=\pi/4, ~\a_0(3)=3/4$, \etc.  Interestingly,
$\a_0(\infty)=1/\sqrt{2}$, which can be understood by noting that in the
limit $d\to\infty$, two arbitrary particles always move orthogonally
with relative velocity $v\sqrt{2}$.  The expression for $\a_0(\infty)$
(\back1) essentially involves the inverse of this factor.
\subsec{(b) Stationary and Moving Species}

Consider now the $d$-dimensional analog of the symmetric 3-velocity
model in which there is a finite initial concentration of stationary
particles of radii $R_0$ and mobile particles of radii $R$ all moving
with speeds $v$.  The appropriate rate equations for the corresponding
concentrations $c_0(t)$ and $c(t)$ are similar to those for the above
impurity limit except that one must account for the influence of
stationary particles on moving particles.  The rate equations become
(compare with Eqs.~(4) with $c_+=c_-$)
$$
\eqalign{
\dot c_0 &= -v\O_{d-1}(R+R_0)c_0c,\cr
\dot c~   &= -v\O_{d-1}(R+R_0)[c_0c+\l_d c^2],\cr}
\eqnoi
$$
where $\l_d=\CF(1)\O_{d-1}(2R)/\O_{d-1}(R+R_0)$.
Introducing the modified time variable,
$T=v\O_{d-1}(R+R_0)\int_0^tdt'\,c(t')$, gives the linear equations
$$
\eqalign{
\td {c_0} T &= -c_0,\cr
\td c T &= -c_0-\l_d c,\cr}
\eqnoi
$$
which can be readily solved to give
$$
\eqalign{
c_0(T) &= c_0(0)e^{-T},\cr
c(T)   &= c(0)e^{-\l_d T} -c_0(0){e^{-T}-e^{-\l_d T}\over \l_d -1}.\cr}
\eqnoi
$$

For $\l_d \ge 1$,
the concentration of the mobile species decays exponentially in real
time $t$ while stationary particles always persist:
$$
c_0(\infty)=c_0(0)\left[1+(\l_d-1){c(0)\over c_0(0)}\right]
^{-1/(\l_d-1)}.
\eqnoi
$$

For $\l_d<1$, there are three different regimes.  For small initial
concentration of mobile species, $c(0)<1/(2-\l_d)$, the density of
moving particles decays exponentially in time, while stationaries
persist with a residue still given by \last.  At the critical
point, $c(0)=1/(2-\l_d)$ and $c_0(0)=(1-\l_d)/(2-\l_d)$, both species
decay as $t^{-1}$. Finally, for $c(0)>1/(2-\l_d)$, both species decay as
distinct power laws in time: $c(t)\sim t^{-1}\gg c_0(t)\sim
t^{-1/\l_d}$.  Amusingly, these same three qualitative regimes occur in
the one-dimensional 3-velocity system.  For example, in three
dimensions, $\l_3=\sqrt{24}({2R\over R+R_0})^2$; hence the threshold
between different behaviors, $\l_3=1$, occurs when $R_0/R\cong 3.42673$.
Thus the stationary particles always persist if their relative size is
small, $R_0/R \le 3.42673$, while for $R_0/R > 3.42673$, stationary
particles may disappear if their initial concentration is small enough.

\subsec{(c) The 2-Speed Model}

For the $d$-dimensional 2-speed model, with $c_j, R_j$, and
$v_j$ the concentration, radius, and speed of \jth\ species, $j=1,2$,
with $\e\equiv v_1/v_2<1$, the rate equations are
$$
\eqalign{
\dot c_1 &= -c_1^2-\m c_1c_2,\cr
\dot c_2 &= - \m c_1c_2 - \n c_2^2,\cr}
\eqnoi
$$
where we have set $v_1\O_{d-1}(2R_1)\CF(1)=1$ by rescaling the
time and have introduced
$$
\m =\e^{-1}{\O_{d-1}(R_1+R_2)\over{\O_{d-1}(2R_1)}}
{\CF(\e)\over{\CF(1)}},\qquad
\n =\e^{-1}{\O_{d-1}(R_2)\over{\O_{d-1}(R_1)}}.
\eqnoi
$$

\back1\ give rise to three asymptotic behaviors which
depend on the relative magnitudes of $\m$ and $\n$:
$$
\eqalign{
c_1(t) & \sim t^{-1},    \qquad c_2(t) \sim t^{-\m},\qquad 1<\n<\m,\cr
c_1(t) & \sim t^{-\m/\n},\quad c_2(t) \sim t^{-1}, \qquad \n<\m<1,\cr
c_1(t) & \sim t^{-1}, \qquad  c_2(t) \sim t^{-1}, \qquad \n<\m<1, \m<1<\n.\cr}
\eqnoi
$$
In the two remaining cases of $1<\n<\m$ and $\n<1<\m$ the asymptotic
behavior depends also on initial concentrations. The first and the
second asymptotics are realized when ${\m-1\over \m-\n}\gl {c_2(0)\over
c_1(0)}$, respectively, while the third asymptotics occurs if
${\m-1\over\m-\n}={c_2(0)\over c_1(0)}$.  Parenthetically,
for an arbitrary number of mobile species
of equal radii in three dimensions, a similar analysis gives
$$
c_j(t) \sim t^{-\m_j},
\quad \m_j={v_1\over 4v_j}+{3v_j\over 4v_1},
\eqnoi
$$
where $v_1$ is the smallest velocity (therefore, $\m_1\equiv 1$).
Thus the least mobile species decays as  $t^{-1}$,
while the more mobile species decay non-universally.

\chap{KINETICS IN ONE DIMENSION}
\tsubsec{(a) Geometric Approach for the Symmetric 2-Velocity Model}

For completeness and to provide a framework to discuss the 3- and
4-velocity models, we first give a geometric derivation for the decay of
the concentration in the 2-velocity or $\pm$ model when the initial
concentrations of the two species are equal.  This approach is based
on the equivalence between the kinetics of the particle system and the
smoothing of one-dimensional stepped interface (Fig.~3).  In this
mapping, a right-moving particle in the $\pm$ model, is equivalent to an
``up'' step in the corresponding interface.  This up step moves to the
right at the same speed as the initial particle.  Similarly, a
left-moving particle is equivalent to a left-moving ``down'' interface
step.  An annihilation event in the particle system corresponds to the
disappearance of a tier in the interface.  Clearly, the collision
partner of a given up step is the first down step to the right which is
at the same height as the initial step.

Since up and down steps steps are uncorrelated and occur with equal
probability, the probability $f_n$ that the initial up step (defined to
be particle 0) collides with its $(2n+1)^{\rm st}$ neighbor is exactly
equal to the first passage probability for a random walk to return to
the origin at $2n+2$ steps.  Thus the initial right-moving particle
annihilates with a left-moving $(2n+1)^{\rm st}$ particle with
probability [7]
$$
f_n = 2^{-2n-1}{(2n)!\over n!(n+1)!}\eqnoi
$$
Asymptotically, this collision probability varies as $n^{-3/2}$ as
$n\to\infty$.  Consequently, the probability that the initial particle
survives potential encounters up to the 2\nth\ neighbor, $S_n\equiv
1-\sum_{n'\leq n}f_{n'}$, leads to a survival probability which decays
as $S_n \sim n^{-1/2}$, or $c(t)\sim t^{-1/2}$ in the continuum limit.

\subsec{(b) The 3-Velocity Model}

For the 3-velocity model, there does not appear to be a similar
geometric construction to help determine the kinetics.  We therefore
resort to qualitative arguments, as well as numerical simulations, to
determine the long-time behavior.  In one dimension, we expect the
kinetics to be different from mean-field predictions because of the
tendency of like velocity particles to cluster, as observed in
space-time graphs (Fig.~1).  For concreteness, we focus on the symmetric
system where $p_+=p_-\equiv p_\pm$ and first investigate whether
stationary particles persist for any value of $p_0$ by numerical
simulations.  Because of subtle crossover effects, a direct molecular
dynamics approach is inadequate to yield accurate results and we
therefore developed a more efficient approach in which all collision
partners and corresponding collision times are identified at the outset.

In this algorithm, stationary or right-moving particles are placed on a
stack (first in, last out) as they are initially created.  When a
left-moving particle is created, its collision partner is determined
immediately, since this partner is necessarily one of the particles from
the already existing stack.  (There is a particular case in which a
negative velocity particle is deposited when the stack is empty.  This
exits the system, since free boundary conditions are employed.  To
ensure that this effect does not give spurious results, only particles
from the middle half of the system are considered).  The determination
of the collision partner of the left-moving particle is accomplished by
straightforward comparisons.  If the uppermost particle on the stack
moves to the right, then it is the collision partner, and the collision
time is recorded.  On the other hand, if the ``last'' particle on the
stack has zero velocity, one must compare the collision time between
this last particle on the stack and the left-moving particle, and the
collision time between the last particle with ``earlier'' outgoing
particles from the stack.  All collision partners and corresponding
collision times are determined, up to and including the collision time
of the initial left-moving particle.  Right-moving or stationary
particles for which the collision time is determined are then removed
from the stack.  From the stored array of collision times one determines
$c_\pm(t)$ and $c_0(t)$ by counting the number of particles of a given
species that survive at that time.

With this method we simulated $5\times 10^5$ particles to $10^5$ time
steps in approximately 30 cpu seconds on a DEC/AXP 3000/400 workstation.
Our numerical results are typically based on 100 realizations at each
initial concentration.  These simulations reveal the following basic
results (Fig.~4): For $p_0< 1/4$, $c_0(t)\sim 1/t$ and $c_\pm(t)\sim
t^{-1/2}$.  The crossover to the asymptotic behavior becomes
progressively more gradual as $p_0\to 1/4$ from below and there is a
substantial time range for which $c_\pm(t)$ and $c_0(t)$ decay at nearly
the same rate before the final asymptotics is reached.  (This is the
primary reason for the erroneously-reported non-universal behavior based
on data from direct and much less extensive molecular dynamics
simulations [1].)~ Exactly at $p_0=1/4$, the data indicate that both
$c_\pm(t)$ and $c_0(t)$ decay as $t^{-\a}$, where extrapolations of
local slopes of neighboring points in figure 4(b) give $\a\cong 0.665$
(Fig.~5).  When the time difference between these points is the factor
$1.2^{21}\approx 46$, a smooth sequence of local exponents is obtained.
However, similar results are obtained for different choices of time
delay factors.  The trend in the local exponents suggests, in fact, that
$\a=2/3$, a result which has now been obtained by an exact calculation
[5].  Additionally, we estimate the relative amplitude
$c_\pm(t)/c_0(t)\cong 1.17$.  For $p_0>1/4$, $c_0(t)$ saturates to a
finite limiting value which appears to be proportional to $(p_0-{1\over
4})^2$, while $c_\pm(t)$ decays faster than a power law in time.  Based
on these results, the phase diagram shown in Fig.~6 is inferred.

The location of the tricritical point, where all three species decay at
the same rate, may be found by the following heuristic argument [8].
Since half the stationary particles react with + particles, the fraction
of + particles available to react with $-$ particles is simply
$p_+-{1\over 2}p_0$.  This is proportional to the number of $+-$
annihilation events per unit length, $P_{+-}$.  Similarly, the relative
number $P_{0-}$ of $0-$ annihilation events per unit length is equal to
${1\over 2}p_0$.  Now we {\it assume} that $P_{+-}/P_{0-}=2$, based on
the expectation that the relative number of annihilation events is
proportional to the relative velocities of the collision partners.
Combining the resulting relation, $p_+-{1\over 2}p_0=p_0$, with the
normalization condition, $2p_++p_0=1$, we find the location of the
tricritical point to be $p_0=1/4, p_+=p_-=3/8$.

It is straightforward to generalize this argument to the asymmetric
velocity distribution $(+v,0,-1)$.  Since the $+ \leftrightarrow
-$ symmetry is now broken, we must consider separately the ratios,
$P_{+-}/P_{0-}$ and $P_{-+}/P_{0+}$.  Following the same considerations
as in the symmetric case, these reaction numbers are
$$
\eqalign{
{P_{+-}\over P_{0-}} &={p_+- vp_0/(1+v)\over {p_0/(1+v)}}
=1+v,\cr
{P_{-+}\over P_{0+}} &={p_--  p_0/(1+v)\over {vp_0/(1+v)}}
={1+v\over v}.\cr}
\eqnoi
$$
Together with the normalization condition,
we find, for the initial concentrations at the tricritical point,
$$
p_0={1\over 4}, \quad
p_+={1\over 4}\left(1+{v\over 1+v}\right), \quad
p_-={1\over 4}\left(1+{1\over 1+v}\right).
\eqnoi
$$

While \last\ involves uncontrolled approximations, especially since
symmetry considerations no longer apply, it is relatively accurate.  For
example, simulations with $v=2$ suggest that the tricritical point is
located at $p_0\approx1/4$ and $p_+\approx0.402$, compared to
$p_+=5/12=0.41\overline{6}$ from \last.  Generally when $v$ varies from
0 to $\infty$, the tricritical point moves along the straight line
parallel to the base of the composition triangle.  At the extreme limits
of $v=0$ and $v=\infty$ the 3-velocity model degenerates to the
2-velocity model for which we know that \last\ is exact.  The available
numerical evidence also supports the apparent general equality
$p_0=1/4$.  Another interesting question in the general asymmetric case
is whether the three coexistence curves really coalesce at one point or
whether there is an open region where all concentrations decay at a
similar rate.  Numerics cannot answer such a question definitively, but
the evidence appears to favor the hypothesis of one single tricritical
point where all three coexistence lines merge.

Numerical results also suggest that near the tricritical point of the
symmetric model, $p_0=1/4, p_\pm=3/8$, the long time kinetics
depends on a single scaling variable $\xi \equiv t\D^3$ where
$\D=p_0-1/4$ is assumed to be small.  The
scaling assumption for the concentrations gives
$$
c_0(t,\D)     \sim t^{-2/3}\Cat, \quad
c_{\pm}(t,\D) \sim t^{-2/3}\Cbt,
\eqnoi
$$
with $\Cax$ and $\Cbx$ finite at $\xi=0$ to reproduce the $t^{-2/3}$
decay at the tricritical point.  This implies that $c_0(t,\D) \sim
\D^2$, as is observed numerically.  For $\D<0$ the scaling predictions
agree with simulations if $\Cax \sim (-\xi)^{-1/3}$ and $\Cbx \sim
(-\xi)^{1/6}$ as $\xi \to -\infty$.  Thus scaling provides the $\D$
dependence of the asymptotic behavior for $\D<0$, namely, $c_0(t)\sim
(t\D)^{-1}$ and $c_\pm(t)\sim (\D/t)^{1/2}$.  Additionally, the crossover
time from tricritical behavior to the final asymptotics scales as
$\D^{-3}$.

Since the concentration of stationary particles saturates to a finite
value for $\D>0$, $\Cax\sim \xi^{2/3}$ as $\xi \to \infty$.  In this
regime, $c_\pm(t)$ will eventually decay rapidly, presumably
exponentially, leading to a similar decay of $\Cbx$ as $\xi\to\infty$.
This has a remarkable consequence for the spatial distribution of
immobile particles.  Since the crossover time is of order $\D^{-3}$,
$c_\pm(t)$ should decay asymptotically as $\exp(-t\D^3)$.  The residual
concentration of the immobile particles, however, goes as $\D^2$.  The
two facts appear contradictory, since the decay law for $c_\pm(t)$
implies that immobile-free intervals of length of order $\D^{-3}$ must
be reasonably frequent.  The resolution of this difficulty is presumably
that immobile particles are distributed on a fractal set on length
scales up to $\D^{-3}$.  A consistent value for the fractal dimension of
this set is $1/3$, since this value indeed implies that there are
$\D^{-1}$ particles in an interval of length $\D^{-3}$.  This
observation has been verified numerically.

\subsec{(c) The Impurity Limit of the 3-Velocity Model}

When the concentration of one species is negligible while the other two
species have equal concentrations, it is possible to determine the
asymptotic survival probability of the ``impurity''.  These results are
worth emphasizing, both for methodological interest and because of their
unusual nature.  First consider a stationary impurity in a background of
$\pm$ particles.  By the mutual annihilation of $\pm$ particles, their
density decays as $c_\pm(t)\sim t^{-1/2}$.  On the other hand, a
stationary particle survives only if it is not annihilated by particles
incident from either direction.  Since each of these two events is
independent, it follows that $c_0(t)\sim c_{\pm}(t)^2\sim t^{-1}$ in the
limit $p_0\ll p_\pm$.  This same argument continues to apply if the
impurities are ``slow'', \ie, their velocity $w$ satisfies $|w|<1$.  The
full survival probability is again a product of single-sided survival
probabilities, each of which decays as $t^{-1/2}$.

More precisely, suppose that the impurity starts at the origin with
velocity $w$ and that there is a Poisson distribution of initial
separations of the background particles.  The impurity survival
probability at time $t, S(t,w)$, equals the product of left- and
right-sided survival probabilities, $S_-(t,w)S_+(t,w)$.  Since
$S_-(t,w)\equiv S_+(t,-w)$, we only need to compute $S_+(t,w)$.  This
one-sided survival probability is given by
$$
S_+(t,w)=\sum_{n=0}^{\infty}f_nP[x_{2n+1}>(1+w)t].
\eqnoi
$$
This is simply the sum over all collisions partners of the
probability that the collision time between the impurity and
each of its potential collision partners is greater than $t$.
For a Poisson distribution of interparticle separations with unit
density, the probability that a left-moving $(2n+1)^{\rm st}$
particle is located at $x_{2n+1}>(1+w)t$ is
$$
P[x_{2n+1}>(1+w)t]=\int_{(1+w)t}^{\infty}{z^{2n}\over (2n)!}e^{-z}dz.
\eqnoi
$$
Combining \backs5, \backn1, and \lastn\ yields [9]
\vskip 0.1truein
$$
\eqalign{
S_+(t,w)&=1-\int_{0}^{(1+w)t}e^{-z}I_1(z){dz\over z},\cr\cr
        &=e^{-(1+w)t}(I_0[(1+w)t]+I_1[(1+w)t]).\cr}
\eqnoi
$$
\vskip 0.02truein
\noindent where $I_1$ is the modified Bessel function.
Thus the complete survival probability is
\vskip -0.1truein
$$
\eqalign{
S(t,w)&=e^{-2t}(I_0[(1+w)t]+I_1[(1+w)t])\times(I_0[(1-w)t]+I_1[(1-w)t]),\cr\cr
      &\sim {2 \over \pi \sqrt{1-w^2}}~t^{-1}.\cr}
\eqnoi
$$
\vskip -0.05truein
\noindent in agreement with the rough argument given above.

Particularly intriguing behavior occurs in the complementary ``fast''
impurity limit with a vanishingly small initial concentration of $+$
impurities in a background of equal concentrations of $-$ and 0
particles.  By a Galilean transformation, this system can be viewed as
an impurity of velocity $+3$ in a $+\,-$ background.  More generally, we
consider the decay of a fast impurity with speed $|w|>1$ in a $+-$
background.  An asymptotic argument suggests that this survival
probability decays slower than an exponential but faster than a power
law in time.

The basis of this argument is to consider a subset of configurations
which gives the dominant contribution to the impurity survival
probability, but which are sufficiently simple to evaluate.  For the
impurity to survive to time $t$, the background $\pm$ particles must
annihilate among themselves up to this time.  On a space-time diagram,
these self-annihilation events appear as a sequence of isosceles
triangles which do not extend to the world line of the impurity.  We
posit that the dominant contribution to the survival probability stems
from a sequence of systematically larger self-annihilation triangles
which just miss the impurity world line (Fig.~7).  From basic geometry,
the base of the \nth\ triangle, $x_n$, equals $x_0\b^n$, with
$\b=(w+1)/(w-1)$.  Here the separation between successive triangles is
neglected, an approximation which is valid as the number (and size) of
the triangles becomes large.  Under this assumption of abutting
triangles, the number of such triangles that comprise the
self-annihilation sequence up to time $t$ is $N \simeq \ln
(t/x_0)/\ln\beta$.

By construction, collision partners in the background define the sides
of an individual self-annihilation triangle which encloses more local
self-annihilation events.  If these collision partners are separated by
a distance $x_n$, the existence probability of the self-annihilation
triangle is $(x_n/x_0)^{-3/2}$ as $n\to\infty$.  The impurity survival
probability $S(t)$ is therefore the product of occurrence probabilities
for the sequence of self-annihilation triangles up to time $t$, leading
to
$$
\eqalign{
S(t)\sim\prod_n^N(2x_0\beta^n)^{-3/2}&\propto
\beta^{-3N^2/4}\cr &\sim \exp(-\ln^2(t/x_0)/{\scriptstyle{4\over
3}}\ln\beta).\cr} \eqnoi
$$
This result should be accurate when the number of self-annihilation
triangles is large, a situation which occurs when $w\gtwid 1$.  In this
limit, it is impractical to directly simulate the large systems needed
to confirm the above prediction.  We have therefore written a program
which tracks {\it only\/} the impurity and the potential collision
partners at any given stage of the reaction.  Particles which are part
of local self-annihilation events need not, and are not considered,
leading to negligible CPU requirements.  With this method we have
simulated $10^7$ realizations of an impurity with $w$ as small as
$1.0002$.  In this case, the mean impurity lifetime is approximately
340, but there are a few configurations with substantially longer
lifetimes.  For $w<1.005$, a plot of the logarithm of the survival
probability versus $z\equiv \ln^2t/\ln\b$ exhibits good data collapse
and linear behavior over a substantial range (Fig.~8).

{}From $S(t)$ given in \last, the mean impurity lifetime,
$\av{t}=\int^\infty S(t')\,dt'$, is readily computed to be proportional
to $({w+1\over w-1})^{2/3}$.  While this is in excellent agreement with
numerical results, the use of the asymptotic form for $S(t)$ even for
$t$ of order unity in the integral for $\av{t}$ has yet to be justified.
One additional amusing feature is that for $w\to 1^+$, the probability
that the impurity annihilates with a right-moving particle vanishes as
$(w-1)^{1/2}$.  When $w=1$, the impurity becomes one of the right-moving
background particles which can only annihilate with left-moving
particles.

\subsec{(d) The 4-Velocity Model}

We focus on the symmetric system with particles of velocities $\pm v_1$
and $\pm v_2$ and relative concentrations $p_1/2$ and $p_2/2$,
respectively.  According to the mean-field description, the more rapid
particles disappear more quickly for any $p_1, p_2>0$.  However, we find
three regimes of behavior in one dimension (Fig.~9).  Depending on
$\e=v_1/v_2$, there is a critical value $p_1^\ast(\e)$ such that for
$p_1<p_1^\ast(\e)$, $c_1(t)\sim t^{-1}$, while $c_2(t) \sim t^{-1/2}$.
The crossover to these asymptotic behaviors sets at progressively later
times as $p_1$ approaches $p_1^\ast(\e)$ from below.  The converse
behavior occurs for $p_1>p_1^\ast(\e)$.  Roughly speaking, in these two
cases the system reduces to the 2-velocity model as the minority species
disappears.  However, when $p_1=p_1^\ast(\e)$ both species decay at the
same rate.  Based on extensive simulations, this decay appears to be a
power law, $t^{-\a}$, with $\a\cong 0.72\pm 0.01$ (Fig.~9).  This value
is obtained by performing a least-squares fit to the data on a double
logarithmic scale in the time range where linear behavior is most
evident, typically for $10^2\ltwid t \ltwid 10^4$.  The error estimate
is based on the variation of the exponent values for systems with $p_1$
within 0.01 of the critical value.

For detecting this power law behavior, direct molecular dynamics is once
again inadequate.  We therefore developed a more efficient algorithm
which can be viewed metaphorically as ``pick up sticks''.  This approach
is applicable for any initial velocity distribution with non-degenerate
collision times.  We first identify the collision times of all
nearest-neighbor pairs and then sort them in ascending order by a
standard $\CO(N\ln N)$ algorithm [10].  Next, these near-neighbor
collision times are sequentially added to the list of true collision
times if a consistency criterion, to be specified below, is satisfied.
At each storage event, the particle pair $(n,n+1)$ associated with the
underlying collision is removed from the system.  Correspondingly, the
collision times associated with $(n-1,n)$ and $(n+1,n+2)$ must be
discarded, while the collision time associated with the new
nearest-neighbor pair $(n-1,n+2)$ is computed.  If any of these three
collision times is smaller than the next near-neighbor collision time in
the previously sorted list, it is necessary to re-sort the current list
of collision times before continuing with the sequential storages of
true collision times.  Since the largest nearest-neighbor collision
times will never be reached before re-sorting is necessary, the sorting
is performed only on a small fraction of the smallest of these collision
times at each stage.  With this method we simulated $2.5\times 10^5$
particles to $10^5$ time steps in of the order of 70 cpu seconds on a
DEC/AXP 3000/400 workstation.  Our results are typically based on 50
realizations at each initial concentration.

The location of the critical point for the symmetric 4-velocity model
may also be estimated by the same approach that was applied for the
3-velocity model.  Let $P_{jj}$ be the number of annihilation events
between $+$ and $-$ particles of type $j$, $j=1,2$, and let $P_{ij}$,
$i\ne j$, be the number of annihilation events between say a + particle
of type $i$ with both $+$ and $-$ particles of type $j$.  There are two
``mass'' conservation laws, $P_{22}+P_{21}\propto p_2$ and
$P_{11}+P_{12}\propto p_1$, as well as the symmetry condition
$P_{21}=P_{12}$.  We further assume that $P_{22}/P_{21}=2$, indicative
of the fact that the relative velocity between $+$ and $-$
$v_2$-particles is equal to $2v_2$ while the average relative velocity
between say a $+v_2$-particle and an arbitrary $\pm v_1$-particle is
equal to $v_2$. Analogously, one also has $P_{11}/P_{12}=2v_1/v_2$.
Combining these relations, the ratio of initial densities at the
critical point is
$$
{p_2\over p_1}= {3\over 1+2v_1/ v_2}
\eqnoi
$$
\last\ reproduces the expected results in two extreme limits of
$v_1=v_2$, where it gives $p_2=p_1$, while for $v_1=0$, the 3-velocity
tricritical point is reproduced.  For two particular cases that were
simulated extensively, \last\ gives $p_2/p_1=1.5$ for $\e=2$ and
$p_2/p_1=1.065$ for $\e=1.1$, while the corresponding estimates from
simulations are 1.6 and 1.20.  Thus while \last\ reproduces the correct
qualitative trend for the location of the tricritical point, it is less
accurate than the corresponding prediction of the 3-velocity model.

\chap{SUMMARY}

Ballistic annihilation when the particle velocities are drawn from
distributions is a problem that appears ripe for further exploration.
There is a rich array of phenomena in which the underlying discreteness
of the particle velocities is crucial.  We have focused on simple and
generic situations to elucidate these features.  The mean-field
description of ballistic annihilation is deceptively simple but leads to
rather complex behavior.  Particular attention was paid to
distinguishing between the case of one-dimensional kinetics (which is
the basis of conventional mean-field theory) and $d$-dimensional
kinetics, in which averages over all particle directions in $d$
dimensions is accounted for.  Since the ``number'' of species is
infinite, it is difficult to imagine that large-scale single-species
heterogeneities could form. The absence of such spatial organization
suggests that the mean-field approximation is applicable when $d>1$.

In one dimension, the relative initial abundance of various species
fundamentally determines which species predominates in the long-time
limit.  For the symmetric 3-velocity model there is tricritical behavior
in which all three species decay as $t^{-2/3}$ when $p_\pm=3/8$ and
$p_0=1/4$.  In the fast impurity limit, the survival probability of the
impurity decays as $\exp(-{\rm const.}\times \ln^2t)$.  For the
4-velocity model, mean-field theory predicts that the faster species
decays with a non-universal power law, while the slower species always
decays as $1/t$.  However, in one dimension, it is again the relative
initial concentrations which determine whether the more rapid or the
slower species can dominate asymptotically.  At the threshold between
these two regimes, the densities of all species appears to decay at
$t^{-\a}$, with $\a\cong 0.72$.

It is important to mention that Piasecki and Droz \ea\ have very
recently developed a powerful analytic method to solve for the kinetics
of one-dimensional of ballistic annihilation models exactly.  In
particular, they obtain the $t^{-2/3}$ decay of the density at the
tricritical point of the symmetric 3-velocity model that we inferred
from simulations.  Their method also appears to be applicable to general
velocity distributions.  However, even in the three velocity model, the
construction of explicit solutions from their formalism is a formidable
task.  Thus it still would be desirable to develop either continuum
approaches or other analytical methods that would provide better
intuitive insights into the intriguing qualitative features of ballistic
annihilation.

\bigskip\centerline{\bf ACKNOWLEDGMENTS}\medskip

We thank E.~Ben-Naim, M.~Bramson, D.~Dhar, L.~Frachebourg, and
I. Ispolatov for helpful discussions and correspondence.  We also thank
L.~Frachebourg for informing us of related work prior to publication.
We gratefully acknowledge ARO grant DAAH04-93-G-0021, NSF grants
INT-8815438 and DMR-9219845, and the Donors of The Petroleum Research
Fund, administered by the American Chemical Society, for partial support
of this research.

\vfill\eject

\def\pra #1 #2 #3 {{\sl Phys.\ Rev.\ A} {\bf #1}, #2 (#3)}
\def\pre #1 #2 #3 {{\sl Phys.\ Rev.\ E} {\bf #1}, #2 (#3)}
\def\prl #1 #2 #3 {{\sl Phys.\ Rev.\ Lett.} {\bf #1}, #2 (#3)}

\bigskip\centerline{\bf REFERENCES}\medskip

\refi E.~Ben-Naim, S.~Redner, and F.~Leyvraz, \prl 70 1890 1993 .

\refi Y.~Elskens and H.~L.~Frisch, \pra 31 3812 1985 ;
      J.~Krug and H.~Spohn, \pra 38 4271 1988 .

\refi A version of ballistic annihilation with a trimodal velocity
distribution was introduced by W.~S.~Sheu, C.~Van den Broeck, and
K~Lindenberg, \pra 43 4401 1991 .  However, the collision rules of this
model were formulated to give behavior similar to that of the 2-velocity
model.

\refi Some aspects of the kinetics of ballistic annihilation with
discrete velocity distributions were reported by S. Redner in {\it
Proceedings of the International Colloquium on Modern Quantum Field
Theory II}, ed.\ G. Mandal (World Scientific, 1994).

\refi J. Piasecki, preprint; M. Droz, P.-L. Rey, L. Frachebourg, and
J. Piasecki, preprint.

\refi P.~L.~Krapivsky, E.~Ben-Naim, and S.~Redner, \pre 50 2474 1994 .

\refi W.~Feller, {\sl An Introduction to Probability Theory and Its
      Applications} (Wiley, New York, 1968).

\refi E.~Ben-Naim, private communication.

\refi M. Abramowitz and I. A. Stegan, {\sl Handbook of Mathematical
Functions} (Dover Publications, Inc., New York, 1964).

\refi W. H. Press, B. P. Flannery, S. A. Teukolsky, and
W. T. Vetterling, {\sl Numerical Recipes} (Cambridge University Press,
Cambridge, 1986).

\vfill\eject

\bigskip\centerline{\bf FIGURE CAPTIONS}\medskip

\rfigi Space-time representation of particle trajectories in the
symmetric 2- and 3-velocity models, where particles move with velocity 0
or $\pm 1$.  Shown are: (a) $p_0=0.0$, $p_\pm=0.50$ and (b) $p_0=0.25$,
$p_\pm=0.375$.

\rfigi ``Phase'' diagram of the symmetric 3-velocity model in mean-field
limit within the relative composition triangle defined by triangular
region $p_++p_-+p_0=1$.  The regions marked by $+$, $-$, and 0 are
phases where only positive, negative, or stationary particles,
respectively, persist in the long time limit.  Along the boundaries
between $+\,0$ or $-\,0$, the concentrations of the two competing
species decay as $t^{-1}$, while the minority species decays as
$t^{-3}$.  The width of the 0 phase region is vanishingly small as
$p_0\to 0$.

\rfigi  Equivalence between the time evolution of
the 2-velocity model and a one dimensional ``wedding cake'' interface.
The collision partner of the initial particle is indicated in both the
particle and interface representations by the small arrows.

\rfigi Representative simulation results for the symmetric 3-velocity in
one dimension.  Shown are double logarithmic plots of $c_+(t)$ (+),
$c_-(t)$ ($\nabla$), and $c_0(t)$ ($\circ$) versus $t$ for: (a)
$p_0=0.10$, $p_\pm=0.45$, and (b) at the tricritical point $p_0=0.25$,
$p_\pm=0.375$.  Each successive data point represents in increase in $t$
by a factor of 1.2.

\rfigi Local exponents for $C(t)\equiv c_+(t)+c_-(t)$ (+) and $c_0(t)$
(circles) versus $1/t$ at the tricritical point.  These exponents are
based on the slopes of points separated by $\tau=1.2^{21}\approx 46$
from the data of figure 4(b).

\rfigi ``Phase'' diagram of the symmetric 3-velocity model in one
dimension.  Along the dashed line, $c_{\pm}(t)\sim t^{-1/2}$, while
$c_{0}(t)\sim t^{-1}$.  At the tricritical point (circle), all species
decay as $t^{-2/3}$.  Along the solid lines, the nature of the decay is
unknown, except very close to the extrema which correspond to the ``fast
impurity'' problem (see text).

\rfigi World line of a fast impurity in a background of equal
concentrations of background $\pm$ particles.  Successive
triangles of self-annihilating background particles are indicated.

\rfigi Plot of the survival probability of a fast impurity.
Shown is $\ln S(t)$ versus $\ln^2t/\ln\b$ for $w-1= 0.0005$,
0.001, and 0.002 (+, $\D$, and $\times$, respectively).

\rfigi Representative simulation results for the symmetric 4-velocity in
one dimension.  Shown are a series of double logarithmic plots of
$c_1(t)$ ($\square$) and $c_2(t)$ ($\circ$) versus $t$ for: (a) - (c)
$v_1=1$ and $v_2=1.1$ for $p_1=0.40$, 0.45, and 0.50, and (d) - (f)
$v_1=1$ and $v_2=2$ for $p_1=0.35$, 0.38, and 0.40.  Each successive
data point represents in increase in $t$ by a factor of 1.5.

\bigskip\bigskip\vfill\eject\bye